\documentclass[twocolumn,showpacs,preprintnumbers,amsmath,amssymb]{revtex4-1}

\usepackage{graphicx,subfigure}
\usepackage{dcolumn}
\usepackage{bm}

\newcommand{\diff}[3][]{\dfrac{\mathrm{d}^{#1}#2}{\mathrm{d}{#3}^{#1}}}

\newcommand{\pdiff}[3][]{\dfrac{\partial^{#1} #2}{\partial {#3}^{#1}}}

\begin{document}

\title{The Dynamics of Liquid Drops Coalescing in the Inertial Regime}

\author{James E. Sprittles}
\email{J.E.Sprittles@warwick.ac.uk} \affiliation{Mathematics Institute, University of Warwick, Coventry, CV4 7AL, UK,}

\author{Yulii D. Shikhmurzaev}
\email{Y.D.Shikhmurzaev@bham.ac.uk} \affiliation{School of
Mathematics, University of Birmingham, Birmingham,  B15 2TT, UK.}

\date{\today}

\begin{abstract}
We examine the dynamics of two coalescing liquid drops in the `inertial regime', where the effects
of viscosity are negligible and the propagation of the bridge front connecting the drops can be considered as `local'. The solution fully computed in the framework of classical fluid-mechanics allows this regime to be identified and the accuracy of the approximating scaling laws proposed to describe the propagation of the bridge to be established.  It is shown that the scaling law known for this regime has a very limited region of accuracy and, as a result, in describing experimental data it has frequently been applied outside its limits of applicability.  The origin of the scaling law's shortcoming appears to be the fact that it accounts for the capillary pressure due only to the longitudinal curvature of the free surface as the driving force for the process. To address this deficiency, the scaling law is extended to account for both the longitudinal and azimuthal curvatures at the bridge front which, fortuitously, still results in an explicit analytic expression for the front's propagation speed.  This new expression is then shown to offer an excellent approximation for both the fully-computed solution and for experimental data from a range of flow configurations for a remarkably large proportion of the coalescence process.  The derived formula allows one to predict the speed at which drops coalesce for the duration of the inertial regime which should be useful for the analysis of experimental data.
\end{abstract}


\maketitle

\section{Introduction}

The rapid motion that ensues after two drops of the same liquid come into contact (Figure 1) is the key element of a
wealth of processes, notably in micro- and nanofluidic devices such as `3D-Printers', where structures are built using microdrops as building blocks. It is clear then, that understanding the physical mechanisms which govern the drops' coalescence, and being able to predict the motion of the drops during this process, is key for the development of these emerging technologies.
\begin{figure}[h]
\begin{minipage}[c]{.49\textwidth}
\subfigure[$\bar{t}=0.04$]{\includegraphics[scale=0.25]{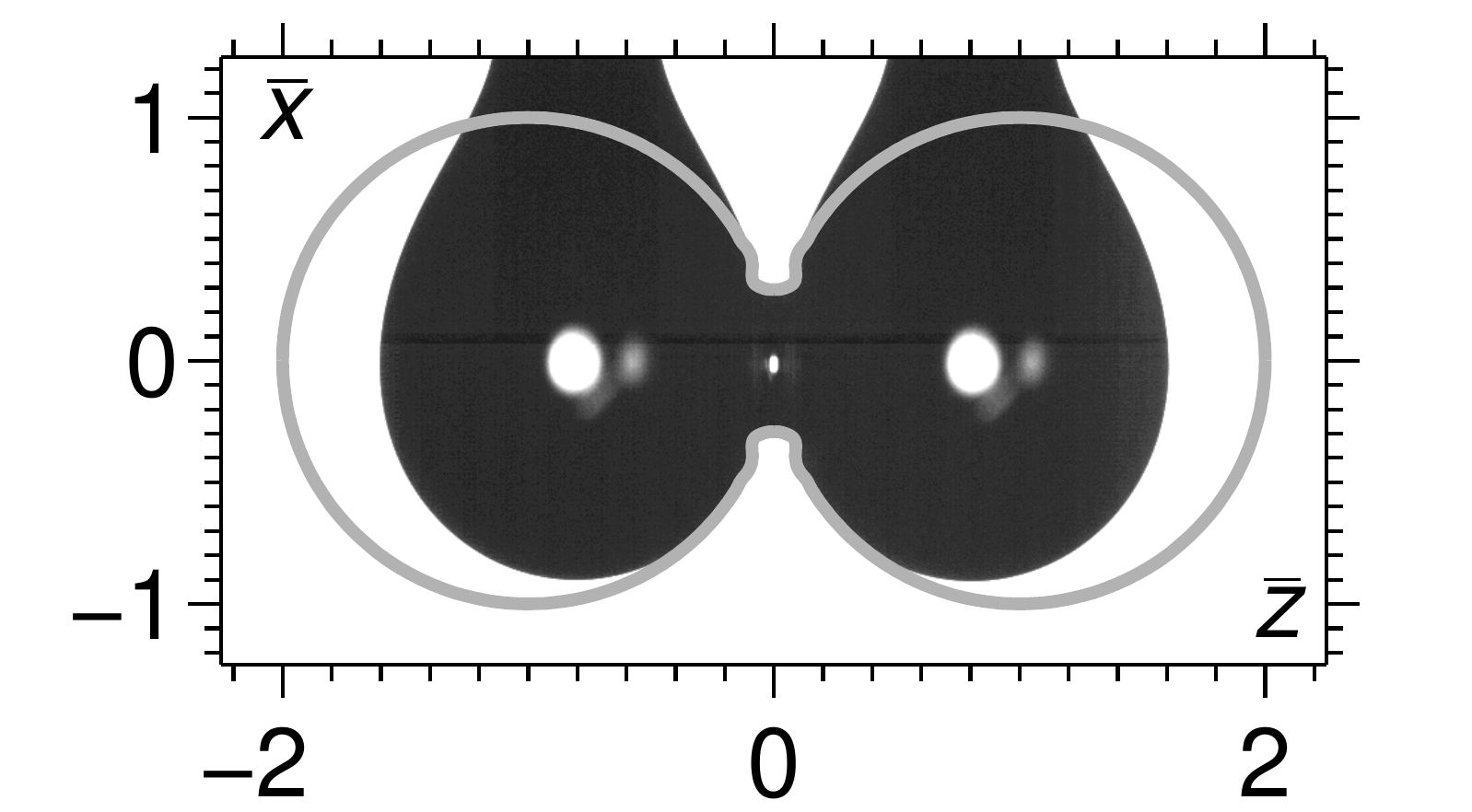}}
\subfigure[$\bar{t}=0.1$]{\includegraphics[scale=0.25]{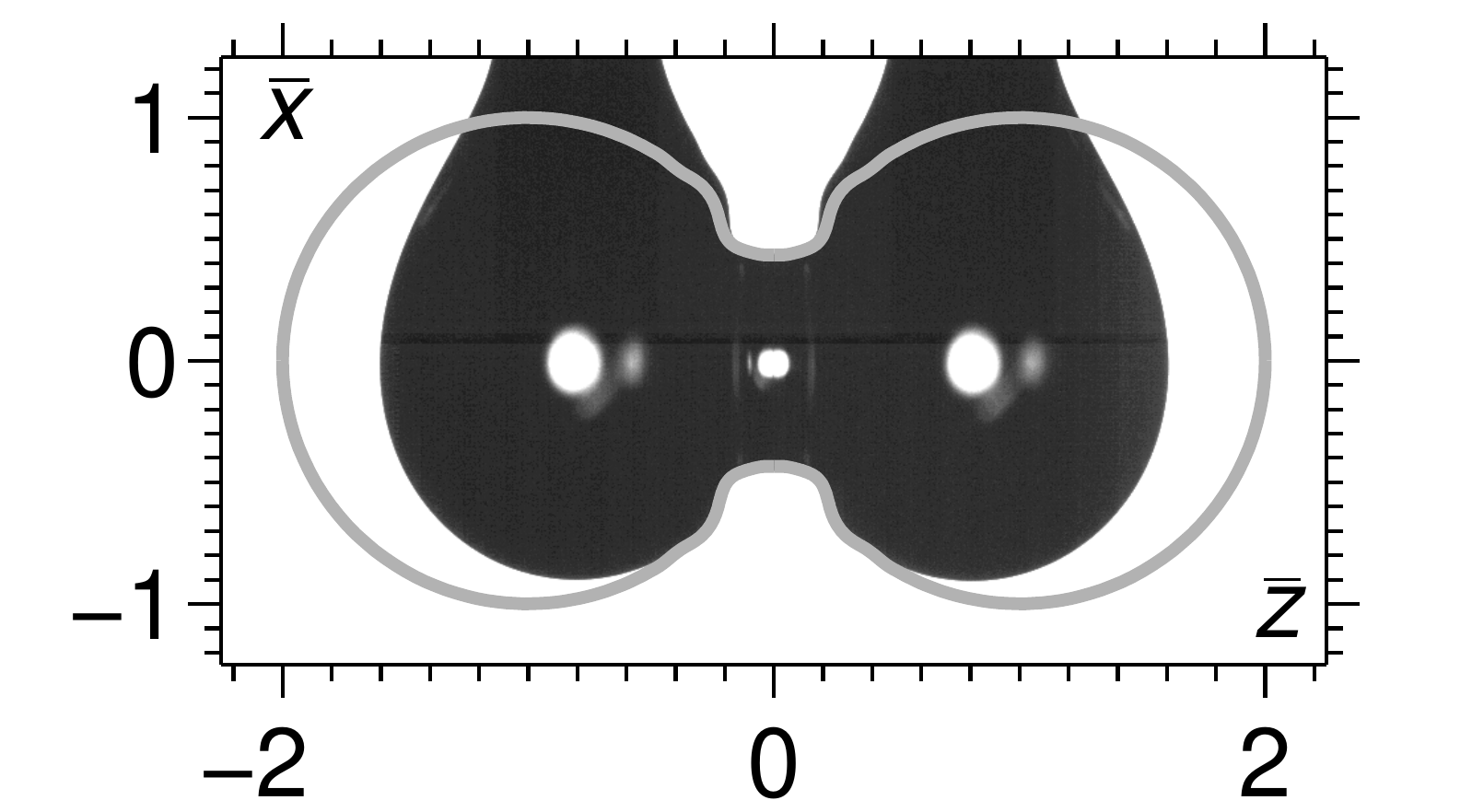}}
    \end{minipage}
     \begin{minipage}[l]{.49\textwidth}
\subfigure[$\bar{t}=0.2$]{\includegraphics[scale=0.25]{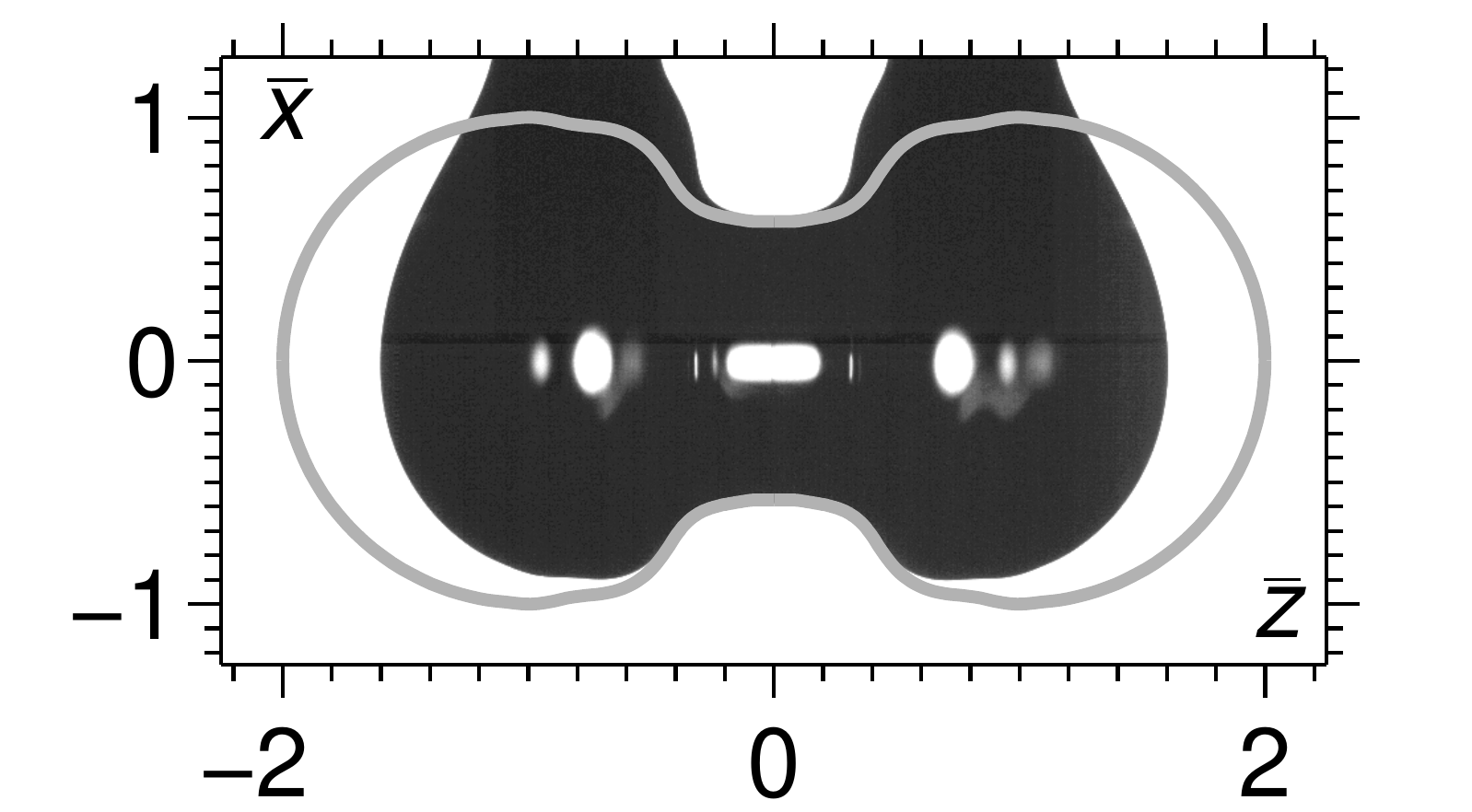}}
\subfigure[$\bar{t}=0.3$]{\includegraphics[scale=0.25]{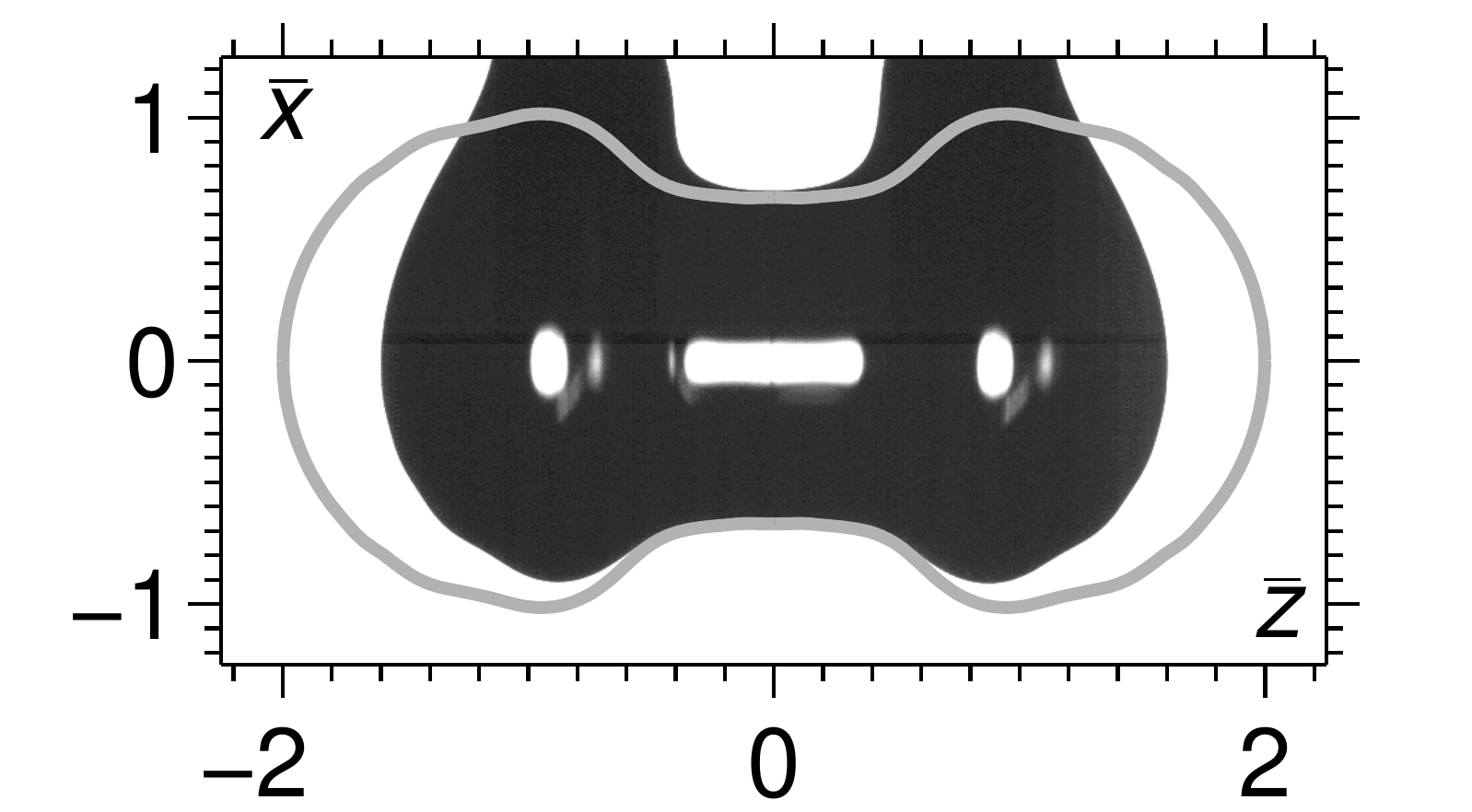}}
    \end{minipage}
 \caption{A typical coalescence event comparing our computations, with free spheres, against experiments in \cite{paulsen12} conducted using 1cP pendent drops of silicone oil (1 unit of length is $R=1.1$~mm and 1 unit of time is $T_i=8$~ms).}\label{F:inertial_pics}
\end{figure}

Due to recent advances in both experimental and computational techniques, there has been a surge in the number of publications studying the coalescence of liquid drops in an ambient gas (air) \cite{menchacarocha01,thoroddsen05,wu04,aarts05,paulsen11,fezzaa08}.  The experimental aspects of the problem have been driven by the application of both ultra high-speed imaging techniques \cite{thoroddsen05} and a novel electrical method \cite{paulsen11}, which has circumvented fundamental issues with optical measurements.  From a computational perspective, specially-designed codes have been used to capture all scales in the problem and to resolve a flow which is known to be singular \cite{sprittles_pof2,paulsen12}.  Notably, most of the aforementioned works have focussed on the different `regimes' encountered and the `transitions' between them, typically shown on log-log plots, with the main attention to formulating or using the correct `scalings' in each regime.

It has now been established that the crossover from the `viscous', or `inertially-limited viscous' \cite{paulsen12}, regime to an `inertial regime', in which viscous forces are negligible compared to inertial ones, occurs when the dimensional radius $r_b$ of the bridge (Figure~\ref{F:sketch}) connecting the coalescing drops in the early stages of the process satisfies $\bar{r}_b=r_b/R \sim \textrm{Re}_i^{-1}$ \cite{paulsen11,sprittles_pof2}, where $\textrm{Re}_i=\sqrt{\rho\sigma R/\mu^2}$ is the Reynolds number in the inertial regime for a drop of radius $R$, density $\rho$, surface tension $\sigma$ and viscosity $\mu$.  This Reynolds number is related to the Ohnesorge number $\textrm{Oh}$ sometimes used in coalescence studies via $\textrm{Oh}=\textrm{Re}_i^{-1}$.

Consider now the bridge radius at which water drops will enter the inertial regime.  If the drops are millimetre-sized $R=1$~mm, as is often the case in experiments, we have $\textrm{Re}_i=O(10^2)$ so that the drops enter the inertial regime when $r_b/R=O(10^{-2})$. If instead microdrops are considered with, say, $R=10~\mu$m, then $\textrm{Re}_i=O(10)$ and the bridge radius still needs only to reach $r_b/R=O(10^{-1})$ before the inertial regime is entered.  In other words, for low-viscosity liquids like water the majority of the dynamics of the bridge (defined, crudely, as $r_b/R>0.1$) of the coalescence event occurs in the inertial regime, even for the drops encountered in microfluidics.  This is the regime which will be considered in this paper.

The inertial regime has previously been studied experimentally, using ultra high-speed cameras \cite{menchacarocha01,thoroddsen05,wu04,aarts05,paulsen12}; analytically, by developing scaling laws \cite{eggers99} and asymptotic theory \cite{thompson12}; and computationally, considering either the local problem \cite{duchemin03}, where the initial stages of bridge front propagation are studied independently from the overall flow configuration, \emph{or} the global dynamics of the drops \cite{menchacarocha01,baroudi14}, where the entire geometry is accounted for.  It has been shown theoretically \cite{eggers99}, computationally \cite{duchemin03,paulsen12,sprittles_pof2} and experimentally \cite{menchacarocha01,thoroddsen05,wu04,aarts05,paulsen12} that, in this regime, the bridge front propagates with a square root in time scaling. In particular, in \cite{eggers99}, the driving capillary pressure $\sigma\kappa$ due to the surface
tension and based on the longitudinal curvature $\kappa\sim 1/d(t)$ obtained from the undisturbed free-surface shape of the drops $d(t)\sim r_b^2(t)/R$ is
balanced by the dynamic pressure $\rho \left(dr_b/dt\right)^2$. As a result, one has $r_b/R = C_{i}\left(t/T_{i}\right)^{1/2}$, where
$C_{i}$ is a constant of proportionality, so that, once non-dimensionalised by our characteristic scales in this regime, that is $R$ for length and $T_i=\sqrt{\rho R^3/\sigma}$ for time,  the scaling law takes the form
\begin{equation}\label{io}
\bar{r}_b=C_{i}\bar{t}^{1/2}.
\end{equation}
Here, and henceforth, all quantities with a `bar' are dimensionless.

Our approach here will be to establish the existence of a well-defined inertial regime, to study the accuracy of scaling laws in this regime using the corresponding numerical solution of the full-scale mathematical problem and to compare their predictions to experimental data from the literature.  This will lead us to an improved scaling law for the regime that will be shown to describe experimental data for a much larger period of time than (\ref{io}) and will allow us to identify previous works where the wrong value of $C_i$ in (\ref{io}) has been chosen.

\section{Problem formulation}

\begin{figure}[h]
     \centering
\includegraphics[scale=0.4]{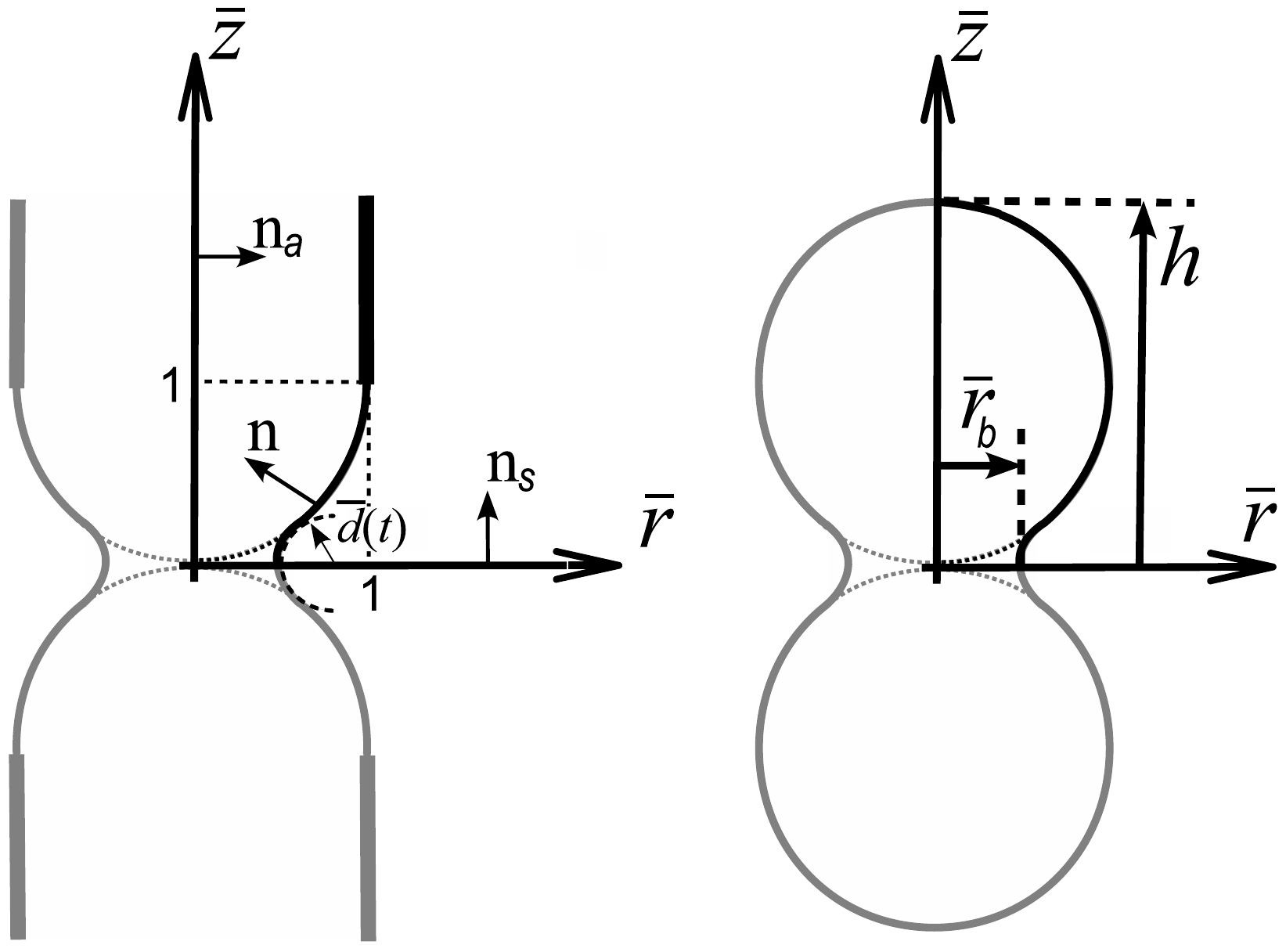}
 \caption{A definition sketch for the coalescence of two identical `pinned hemispheres' grown from syringes (left) and a sketch of coalescing `free spheres' (right) showing the bridge radius $\bar{r}_b$ and apex height $h$.}
 \label{F:sketch}
\end{figure}

In this work we will consider both the typical experimental setup in which hemispherical drops are grown from syringes as well as the case of most practical interest, where free spheres coalesce (Figure~\ref{F:sketch}).  Assuming that gravity can be ignored, which is reasonable for mm-sized drops and below \citep{sprittles_pof2} the
problem becomes symmetric and can be reduced to determining the motion of one drop in the $(\bar{r},\bar{z})$-plane of a cylindrical coordinate system with the symmetry conditions on the $\bar{z}=0$ plane at which the drops initially touch. The syringe, when considered, is taken to be a
semi-infinite cylinder with zero-thickness walls located at
$\bar{r}=1,\bar{z}>1$. The precise far field conditions, i.e.\ those associated with the syringe head,
have a negligible effect on the initial stages of coalescence
\citep{sprittles_pof2}.

To non-dimensionalise the system of the
governing equations for the bulk variables, we use the drop radius
$R$ as the characteristic length scale, $U_{i}=\sqrt{\sigma/(\rho R)}$ as the scale for
velocities, $T_{i}=\sqrt{\rho R^3/\sigma}$ as the time scale and $\mu U_i/R$ as the scale
for pressure. Then, the continuity and momentum balance equations take the form
\begin{gather}\label{ns}
\nabla\cdot\mathbf{u} = 0,\qquad \textrm{Re}_i~\left[\pdiff{\mathbf{u}}{t} + \mathbf{u}\cdot\nabla\mathbf{u}\right] = \nabla\cdot\mathbf{P};\\
\notag\mathbf{P} = -p\mathbf{I} + \left[\nabla\mathbf{u}+
 \left(\nabla\mathbf{u}\right)^T\right],
\end{gather}
where $\mathbf{P}$, $\mathbf{u}$ and $p$ are the (dimensionless) stress
tensor, velocity and pressure in the fluid; $\mathbf{I}$ is
the metric tensor of the coordinate system. The Reynolds number is $\textrm{Re}_i=\sqrt{\rho \sigma R/\mu^2}$.

The conventional boundary conditions used for free-surface flows are
the kinematic condition, stating that the fluid particles forming
the free surface stay on the free surface at all time and
the balance of tangential and normal forces acting on an element of
the free surface from the two bulk phases and from the neighbouring
surface elements:
\begin{equation}\label{ckin}
\pdiff{f}{\bar{t}} + \mathbf{u}\cdot\nabla f = 0
\end{equation}
\begin{equation}\label{cstress}
 \mathbf{n}\cdot\mathbf{P}\cdot\left(\mathbf{I}-\mathbf{n}\mathbf{n}\right) =\mathbf{0},
  \qquad
 \mathbf{n}\cdot\mathbf{P}\cdot\mathbf{n} =\textrm{Re}_i \nabla\cdot\mathbf{n}.
\end{equation}
Here $f(\bar{r},\bar{z},\bar{t})=0$ describes the \emph{a priori} unknown free-surface
shape, with the  unit normal vector $\mathbf{n} = \nabla f/|\nabla
f|$ pointing into the liquid, and the tensor
$(\mathbf{I}-\mathbf{n}\mathbf{n})$ extracts the component of a
vector parallel to the surface with the normal $\mathbf{n}$.

At the plane of symmetry $\bar{z}=0$, the standard symmetry conditions of
impermeability and zero tangential stress are applied
\begin{eqnarray}\label{csym}
\mathbf{u}\cdot\mathbf{n}_s = 0, \qquad \mathbf{n}_s\cdot\mathbf{P}\cdot \left(\mathbf{I}-\mathbf{n}_s\mathbf{n}_s\right) =\mathbf{0}, 
\end{eqnarray}
where $\mathbf{n}_s$ is the unit normal to the plane of symmetry. In
the conventional model we are studying here, the free surface is assumed to always be
smooth so that where it meets the plane of symmetry we have
$\mathbf{n}\cdot\mathbf{n}_s=0$.

On the axis of symmetry $\bar{r}=0$, the standard normal and tangential
velocity condition state that the velocity has only the component
parallel to the $\bar{z}$-axis and the radial derivative of this component is
zero (the velocity field is smooth at the axis),
\begin{eqnarray}
\mathbf{u}\cdot\mathbf{n}_a = 0,
 \quad
 \frac{\partial}{\partial \bar{r}}
 [ \mathbf{u} \cdot (\mathbf{I}-\mathbf{n}_a\mathbf{n}_a) ] =0,
 \qquad & \bar{r}=0;
\end{eqnarray}
where $\mathbf{n}_a$ is the unit normal to the axis of symmetry in
the $(\bar{r},\bar{z})$-plane.

For the case of coalescing free spheres, the free surface is assumed smooth at the apex $\bar{r}=0, \bar{z}=h(\bar{t})$ so that $\mathbf{n}\cdot\mathbf{n}_a=0$ there, whilst the case of coalescing pinned hemispheres requires more conditions to account for the presence of the syringe.  Specifically, at the point in the $(\bar{r},\bar{z})$-plane where the
(initially hemispherical) free surface meets the syringe tip, we have
a pinned contact-line:
\begin{equation}\label{pinned_shape}
 f(1,1,\bar{t})=0 \qquad (\bar{t}\ge0).
\end{equation}

It is assumed that in the far field, the liquid
inside the syringe are at rest, so that
\begin{equation}\label{pinned_shape}
\mathbf{u}\rightarrow\mathbf{0} \qquad\hbox{as}\qquad \bar{r}^2+\bar{z}^2\rightarrow\infty,
\end{equation}
whilst on the cylinder's surface, no-slip is applied
\begin{equation}\label{pinned_shape}
\mathbf{u}=\mathbf{0} \qquad\hbox{at}\qquad \bar{r}=1,\bar{z}\ge 1.
\end{equation}

Computations are started from a finite initial bridge radius $\bar{r}_{min}$, and the details of the initial conditions can be very important when considering the initial stages of motion \cite{sprittles_pof2}.  However, when considering the global motion of the drops, so long as $\bar{r}_{min}$ is sufficiently small, say $\bar{r}_{min}<10^{-2}$, the subtleties surrounding the implementation of the initial conditions are unimportant.  Our computations are started from $\bar{r}_{min}=10^{-4}$ and as an initial condition for the free-surface we take a shape which provides a smooth free-surface at $\bar{r}=\bar{r}_{min}$ (which a truncated sphere would not) whilst far away from the origin (i.e.\ from the point of the initial contact) it is initially the undisturbed hemispherical/spherical drop. A shape which satisfies these criteria can be taken from \cite{hopper84}, i.e. the analytic two-dimensional solution to the problem for Stokes flow. In parametric form, the initial free-surface shape is taken to be
\begin{eqnarray} \notag
 \bar{r}(\theta) &=& \sqrt{2}(1+m)H\cos\theta, \\  \label{ic2}
 \bar{z}(\theta) &=& \sqrt{2}(1-m)H\sin\theta,\\\notag
 H&=&\left[(1-m^2)(1+m^2)^{-1/2}(1+2m\cos\left(2\theta\right)+m^2)^{-1}\right]
\end{eqnarray}
for $0<\theta<\theta_u$, where $m$ is chosen such that $\bar{r}(0)=\bar{r}_{min}$
is the initial bridge radius, which we choose, and $\theta_u$ is
chosen such that $\bar{r}(\theta_u)=\bar{z}(\theta_u)=1$ for hemispherical drops and $\bar{r}(\theta_u)=0$ for spherical ones. Notably, for
$\bar{r}_{min}\to0$ we have $m\to1$  and $\bar{r}^2+(\bar{z}-1)^2=1$, i.e.\ the drop's
profile is a semicircle of unit radius which touches the plane of
symmetry at the origin as required.

Finally, we assume that the fluid starts from rest:
\begin{equation}\label{ic-u}
 \mathbf{u}=\mathbf{0}\qquad\hbox{at}\qquad\bar{t}=0.
\end{equation}

\section{Computational approach}\label{S:comp}

The coalescence phenomenon requires the solution of a free-boundary problem with effects of
viscosity, inertia and capillarity all present, so that a
computational approach is unavoidable.  To do so, we use a
finite-element framework which was originally developed for dynamic
wetting flows and has been thoroughly tested in
\cite{sprittles_ijnmf,sprittles_jcp} as well as being applied to
flows undergoing high free-surface deformation in
\cite{sprittles_pof}, namely microdrop impact onto and
spreading over a solid surface. Notably, the method implemented in our computational platform has been
specifically designed for multiscale flows, so that the very small
length scales associated with the early stages of coalescence can be
captured alongside the global dynamics of the two drops' behaviour.
In other words, all of the spatio-temporal scales present in electrical measurements \citep{paulsen11},
as well as the scales associated with later stages of the drop's
evolution, which are of interest here, can, for the
first time, be simultaneously resolved. A user-friendly step-by-step
guide to the implementation of the method can be found in
\citep{sprittles_ijnmf,sprittles_jcp} whilst benchmark coalescence simulations are provided in \cite{sprittles_pof2}.

\section{Results}\label{S:inert}

\subsection{Identifying an inertial regime}

\begin{figure}
     \centering
\includegraphics[scale=0.255]{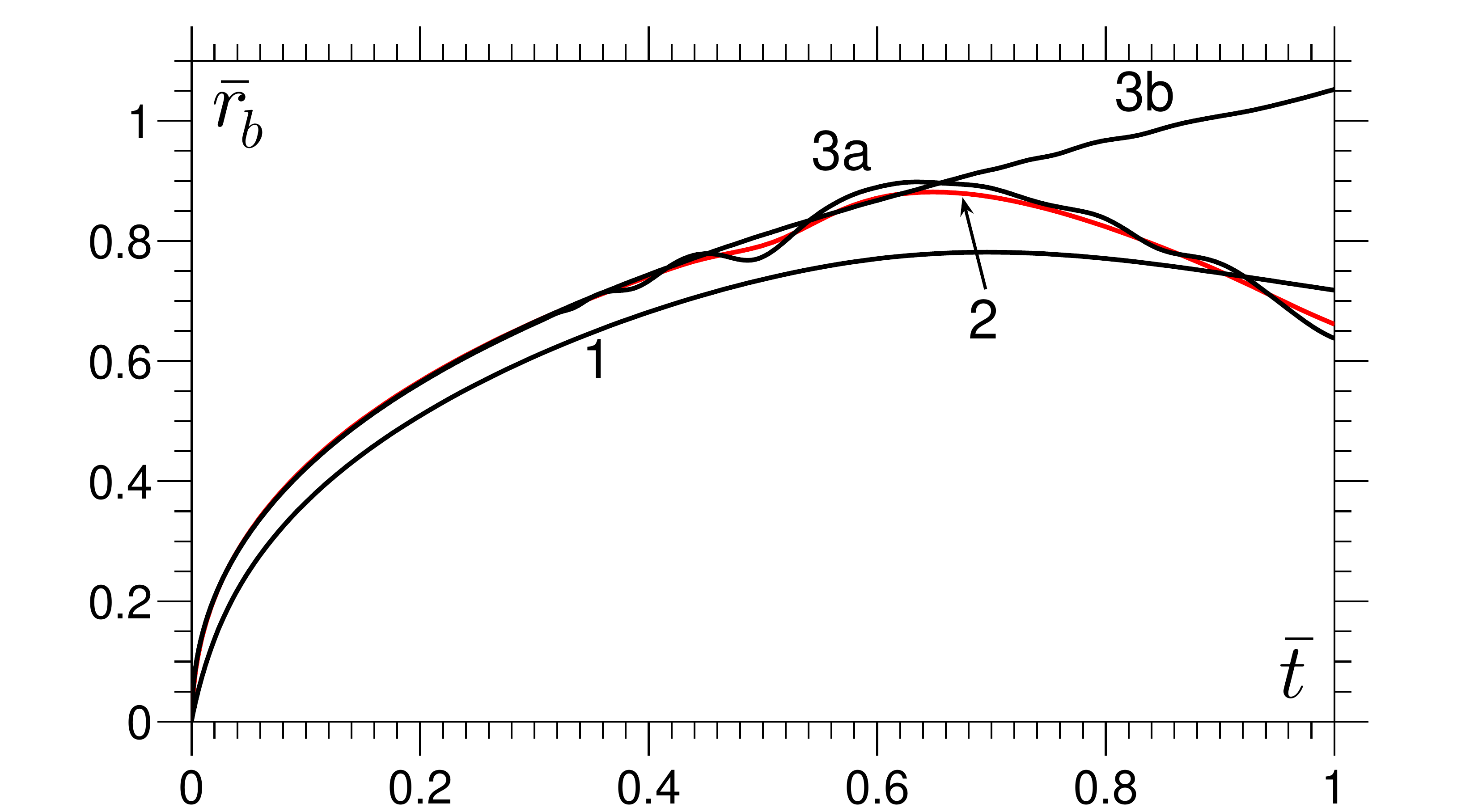}
 \caption{(Color online) Identification of the `inertial regime' showing that
 above a critical Reynolds number, there is a period in which
 the bridge's evolution is independent of the value of $\textrm{Re}_i$
or of the flow configuration.
  Pinned hemispheres are used in computations for 1: $\textrm{Re}_i=10$,
  2: $\textrm{Re}_i=10^2$ (in red), 3a: $\textrm{Re}_i=10^3$,
  whilst curve~3b is for free sphere with $\textrm{Re}_i=10^3$.}
 \label{F:inertial}
\end{figure}

In Figure~\ref{F:inertial}, our computed solutions show that for $\textrm{Re}_i\geq10^2$ (curves 2, 3a and 3b), the evolution of the
bridge between the drops becomes insensitive to further increase in
the Reynolds number until the dimensionless radius of the bridge $\bar{r}_b\approx0.75$. Deviations for very small bridge radii $\bar{r}_b<\textrm{Re}_{i}^{-1}\leq 10^{-2}$, caused by viscous forces being non-negligible and usually observed on a log-log plot, will not be important in the regime we are focusing on.  In this regime hemispheres pinned to the rim of the syringe needles and free spheres (curves 3a and 3b, respectively) give the same results \footnote{It is the capillary waves, initiated at the onset of coalescence, travelling along the free surface and interacting with the far field boundary that eventually `kill' the local nature of the solution in this regime.}.  This suggests that we are truly in an `inertial regime' in
which (a) the effects of viscosity are negligible and (b) the process can be
considered as `local', i.e.\ independent of the far-field
geometry. It is in this regime that scaling law (\ref{io}) is expected to approximate the exact solution and has often
been used to interpret experimental data
\citep{wu04,aarts05,paulsen11}.

\subsection{Standard scalings} 

\begin{figure}
     \centering
\includegraphics[scale=0.255]{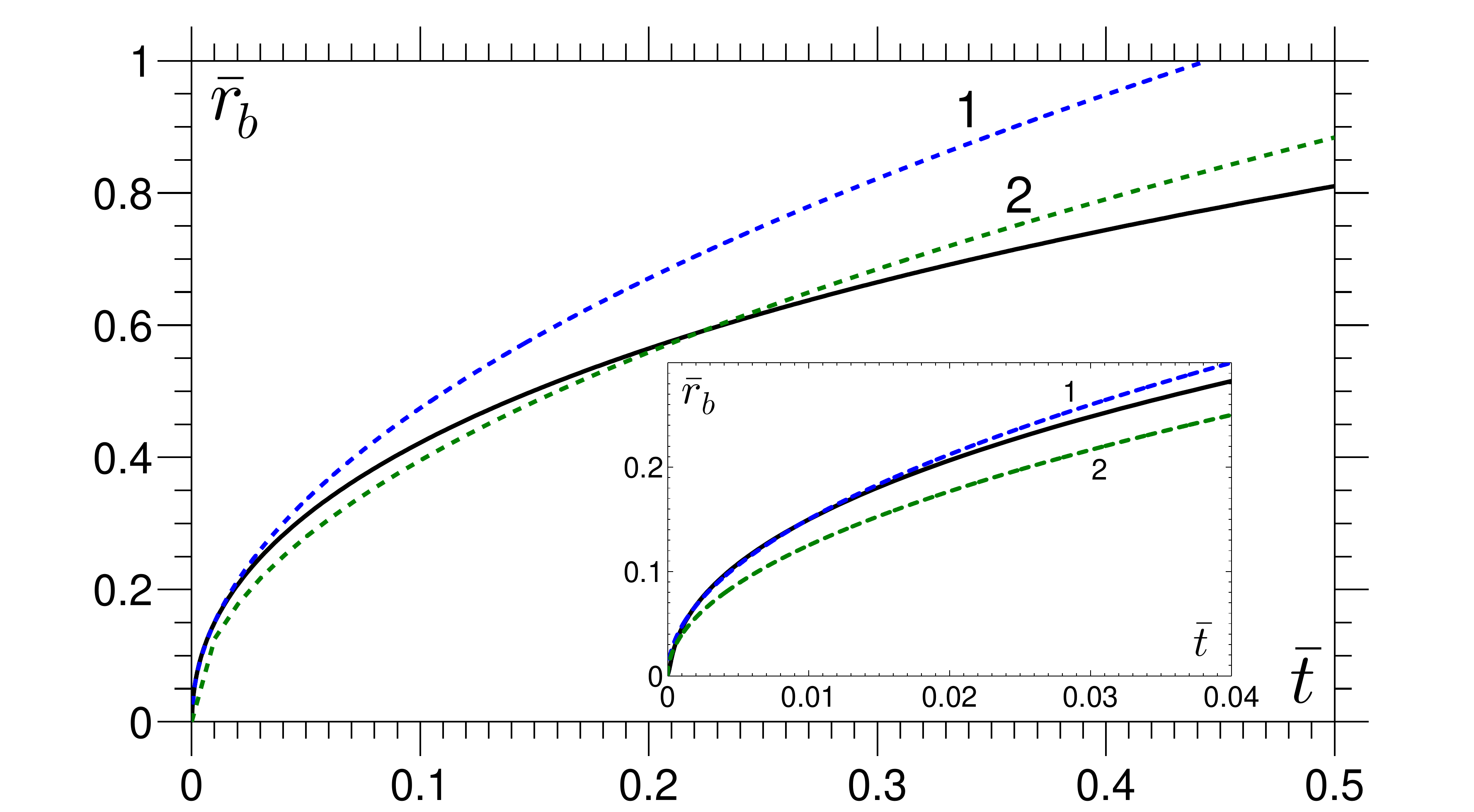}
 \caption{(Color online)  Comparison of the computed solution (solid line) to equation
(\ref{io}) with $C_{i}=1.5$ (curve 1 in blue) and $C_{i}=1.25$ (curve 2 in green).}
 \label{F:inertia_comp}
\end{figure}

In Figure~\ref{F:inertia_comp}, we can see that scaling law
(\ref{io}) with $C_{i}=1.5$ (curve 1) provides a good approximation of the
computed solution (solid line) for $\bar{r}_b<0.15$. However, the scaling law
quickly begins to overshoot the numerical solution. Worryingly, most
comparisons between this scaling law and experimental data has been
in the range accessible to optical observation $\bar{r}_b>0.1$ (which for a
millimetre-sized drop is $r_b=100$~$\mu m$) where our computations show that the scaling law
greatly overshoots the computed solution. In other words, it has
been used \emph{outside} the region where it gives a reasonable
approximation of the solution of the mathematical problem it is
supposed to mimic.

If instead we had looked to `fit' the whole of the computed curve in
the region $0.1<\bar{r}_b<0.75$ as best as we can, ignoring large errors for
$\bar{r}_b<0.1$, then the result is that the prefactor must have a much
smaller value of $C_i=1.25$ (curve 2 in Figure~\ref{F:inertia_comp}) the value which is closer to those
obtained in previous experimental works \citep{aarts05,wu04} that
fitted (\ref{io}).

The upshot of the discrepancy between the scaling law and the
computed solution in the experimental
range is that reported values of $C_i$ have been too small. Although a value of $C_i=1.25$, consistent with those obtained
from experimental analysis \cite{aarts05,wu04}, provides a `best fit' (curve 2 in Figure~\ref{F:inertia_comp})
for $0.1<\bar{r}_b<0.75$ to the exact solution (solid line), and hence also to the
experimental data (Figure~\ref{F:inertia_exp}), as can be seen from Figure~\ref{F:inertia_comp} this solution
completely fails to capture the correct behaviour as $\bar{r}_b\to0$, where
the scaling law should asymptotically approach the exact solution
(solid curve).

\begin{figure}
     \centering
\includegraphics[scale=0.255]{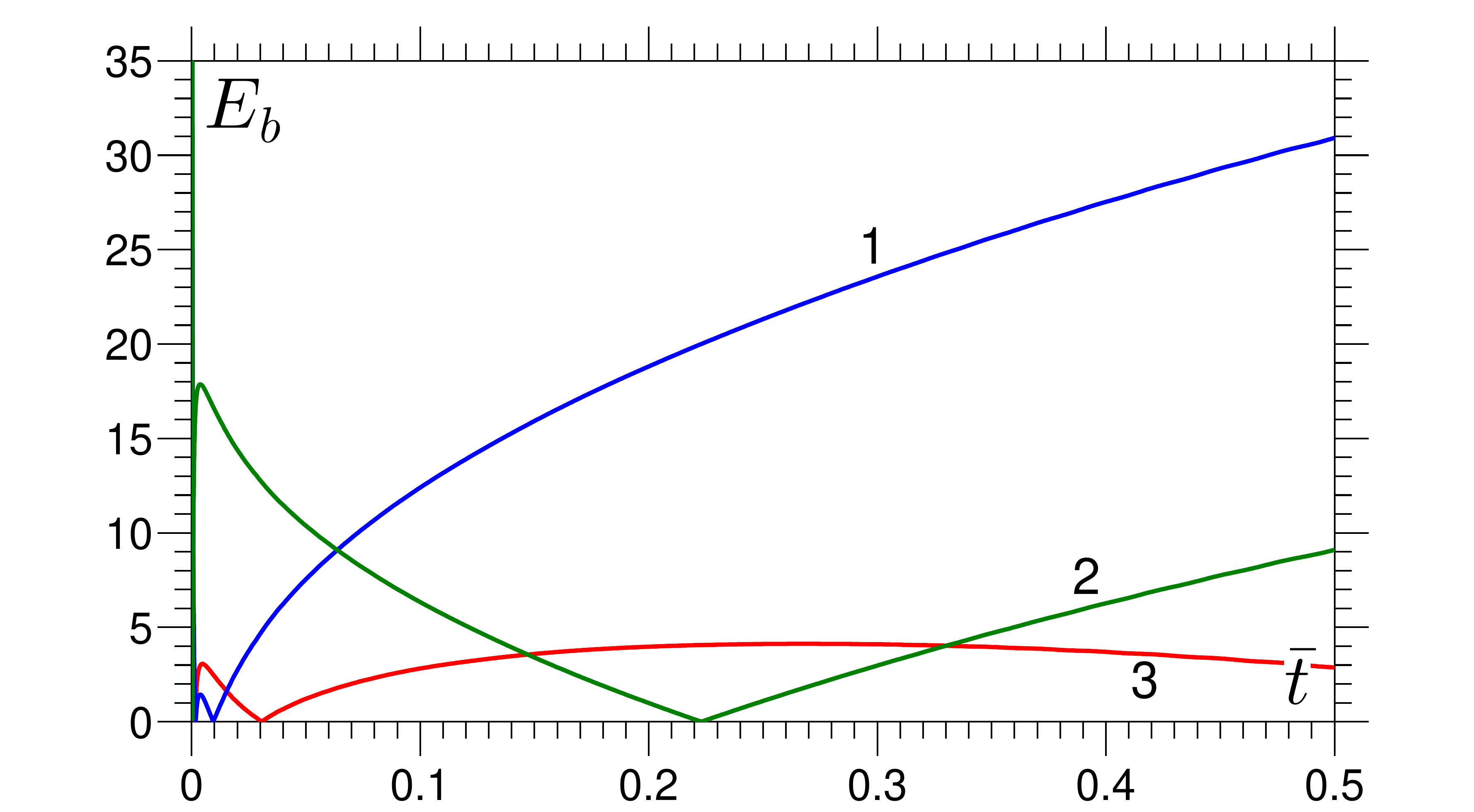}
 \caption{(Color online)  The relative percentage error $E_b$ of the scaling laws from the computed solution. Curve 1 (in blue) is for (\ref{io}) with $C_{i}=1.5$, curve 2 (in green) is (\ref{io}) taking $C_{i}=1.25$ and curve 3 (in red) is (\ref{new_eq3}) with $C_i=1.5$.}
 \label{F:error}
\end{figure}

The failure of the `best fit' approach is confirmed in Figure~\ref{F:error}, where the relative percentage error $E_b(\bar{t})=100|\bar{r}_{sc}-\bar{r}_{co}|/\bar{r}_{co}$ of the scaling laws $\bar{r}_b=\bar{r}_{sc}(\bar{t})$ from the computed solution $\bar{r}_b=\bar{r}_{co}(\bar{t})$ is plotted as a function of time.  One can see that the scaling law (\ref{io}) with $C_i=1.5$ approximates the computed solution well, with $E_b<3\%$ for $\bar{t}<2\times10^{-2}$ whilst during the same time period $E_b>15\%$ for curve 2 which is the `best fit' attempt $C_i=1.25$ in (\ref{io}). The error also confirms that whilst (\ref{io}) with $C_i=1.5$ captures the correct behaviour in the inertial regime for small times, this solution rapidly departs from the computed solution, with $E_b>12\%$ for $\bar{t}>0.1$.  If, as a crude estimate, we require a scaling law to satisfy $E_b<5\%$, then we see that curve 1 meets this criterion for $\bar{t}<0.03$ whilst curve 2 fails in the initial stages and is only valid for $0.12<\bar{t}<0.36$.  Thus, neither of the current scaling laws provide satisfactory approximations to the computed solutions which could be used for a quick comparison between experimental and theoretical predictions.

Here, we will look to rectify the aforementioned inconsistencies by extending the scaling law (\ref{io}), using the approach initiated in \cite{thoroddsen05}, to account for the azimuthal curvature, which reduces the capillary pressure and hence acts
to slow down the evolution of the bridge, as well as  the longitudinal curvature
that drives the process. Although, as we shall see, the latter
dominates as $\bar{r}_b\to0$, it is anticipated that, by including the
azimuthal curvature in the scaling law, we will be able to increase
the region of applicability of our scaling to within the optical range.
This should give a more accurate representation of the
bridge evolution in the inertial regime and can be used to predict the speed of coalescence without having to
resort to computations.

\subsection{An improved scaling} 

Including the curvature at the bridge front in the azimuthal
direction, i.e.\ $\bar{\kappa}_2=-1/\bar{r}_b$, which acts to resist the bridge's outward
motion, into our expression for the full curvature $\bar{\kappa}$ is simple;
however, as a consequence of this extension, we must specify how the
longitudinal curvature $\bar{\kappa}_1$ scales as the bridge propagates since we are no longer able to `absorb' this scale into the constant of proportionality.  Thus, we
now have
\begin{equation}\label{kappa}
\bar{\kappa} = \bar{\kappa}_1 + \bar{\kappa}_2 = \frac{A}{\bar{r}_b^2} - \frac{1}{\bar{r}_b}
\end{equation}
where the constant $A$ must be specified to account for the
longitudinal curvature behaviour as the bridge expands. Previously,
i.e.\ in (\ref{io}), the second term on the right-hand side was neglected and this constant was simply absorbed into $C_{i}$.
If the undisturbed free surface height at $\bar{r}_b$ is taken as the radius
of curvature at that point, then, for small $\bar{r}_b$, we have $A=2$.
However, in \cite{thoroddsen05}, it is argued that $A\approx1$ gives a
better agreement with experiments as with this value the radius of
curvature is the distance between the two undisturbed free surfaces
rather than the distance from the plane of symmetry to one of them
and hence accounts for the ``bulb which develops at the end
of the advancing interface''. This assertion is confirmed by our computations shown in Appendix and so, henceforth, we assume $A=1$
and, if needs be, can later consider whether more accurate representations of $A$ are required. In \cite{thoroddsen05}, the resulting equations, which considered drops
of different sizes, were solved using a numerical method and seen to
give good agreement with the experimental data.

In the case of the coalescence of two identical liquid drops, with
curvature given by (\ref{kappa}), it will be shown that an analytic solution can be obtained which, now $A$ has been specified, still
contains only one free constant.  As proposed in
\cite{eggers99}, balancing (dimensionless) inertial forces with the
(driving) surface tension force gives
\begin{equation}\label{new_eq}
\left(\diff{\bar{r}_b}{\bar{t}}\right)^2 = \frac{C_i^4}{4}\left(\frac{1}{\bar{r}_b^2}-\frac{1}{\bar{r}_b}\right).
\end{equation}
where the coefficient of proportionality has been chosen so that if
the azimuthal curvature is ignored, we recover $\bar{r}_b=C_i\bar{t}^{1/2}$.
Integrating (\ref{new_eq}), assuming that $\bar{r}_b=0$ at $\bar{t}=0$ \footnote{A finite radius of contact could easily be added, but experiments
demonstrate that this radius is very small, so that its inclusion would have almost no effect on the resulting evolution in the regime of interest.}, and
rearranging we obtain a cubic polynomial in $\bar{r}_b$ with $\bar{t}$ as a
parameter:
\begin{equation}\label{new_eq2}
\bar{r}_b^3 + 3\bar{r}_b^2 + \frac{3C_i^2\bar{t}}{4}\left(\frac{3C_i^2\bar{t}}{4}-4\right)=0.
\end{equation}
We can see immediately that if only the leading order terms
in $\bar{r}_b$ and $\bar{t}$ are kept, we have $3\bar{r}_b^2 - 3C_i^2\bar{t}=0$ so that the
usual scaling (\ref{io}) is recovered.  The solution which we require is
given by
\begin{gather}
\notag \bar{r}_b = \frac{s}{4} + \frac{4}{s} - 1, \qquad s = \left[-64-18d^2+96d+2\left(2880d^2\right.\right. \\
\left.\left.-3072d+81d^4-864d^3\right)^{1/2}\right]^{1/3}, \qquad d=C_i^{2}\bar{t},\label{new_eq3}
\end{gather}
where we take the root with the positive imaginary part for $s$ \footnote{This is the default root for most software, e.g.\ MATLAB}, which results in $\bar{r}_b$ being real.

\subsection{Comparison of the improved scaling law to simulations and experiments} 

\begin{figure}
     \centering
\includegraphics[scale=0.255]{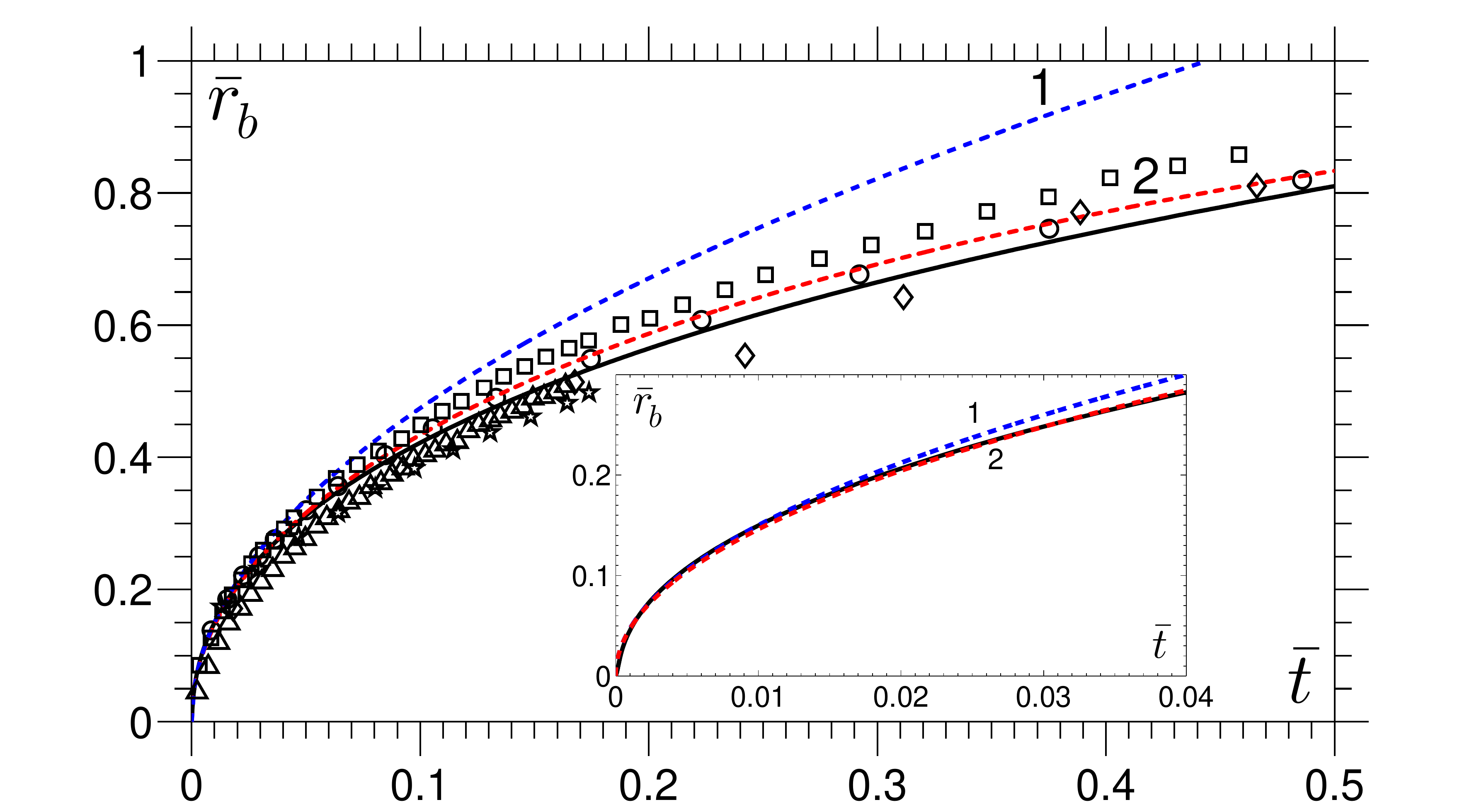}
 \caption{(Color online)  Comparison of the computed solution (solid black curve) of free spheres coalescing with $\textrm{Re}_i=10^3$ to equation
(\ref{io}) with $C_{i}=1.5$ (curve 1 in blue) and the new scaling (\ref{new_eq3}) also with $C_{i}=1.5$ (curve 2 in red). Data has been obtained from the lowest viscosity drops considered in the following publications: circles, Figure~3F in \cite{paulsen12}; squares, Figure~6 in \cite{thoroddsen05}; triangles, Figure~3a in \cite{aarts05}; diamonds, Figure~6 in \cite{menchacarocha01}; and stars, best-fit to Figure~4 in \cite{wu04}.}
 \label{F:inertia_exp}
\end{figure}

The explicit form of (\ref{new_eq3}) allows for quick comparison
with computed solution or experiments, with no additional fitting parameters, in order to determine whether
this is a significant improvement on (\ref{io}).  In the subplot of Figure~\ref{F:inertia_exp}, it can be clearly seen
that for the same value of $C_i=1.5$, the new scaling law (curve 2) gives
results indistinguishable from those given by (\ref{io}) (curve 1) for $\bar{r}_b<0.15$
but for $0.15<\bar{r}_b<0.8$ (shown in the main plot), i.e.\ for the range usually used in
experimental works to fit a scaling law to the data, the new
expression agrees far better with the simulation of the full system (solid line computed for free-spheres)
than the expression (\ref{io}).  This is confirmed in Figure~\ref{F:error} where it can be seen that the new scaling law is within 5\% of the computed solution at least up to $\bar{t}=0.5$. Although curve 2 is not indistinguishable from the numerical result for $\bar{r}_b>0.4$ in Figure~\ref{F:inertia_exp}, as one may expect from the simplifications
made, as an approximating formula it is a significant improvement
on the previous result and is close to the numerical solution for a remarkably large proportion of the coalescence process.

In Figure~\ref{F:inertia_exp}, it can be clearly seen that the new `universal curve' (curve 2) is able to capture experimental
data for the inertial regime collected from the literature
completely different geometric configurations (hanging pendent drops, pinned hemispheres and spheres supported by a hydrophobic solid) and thus provides the
sought-after extension to (\ref{io}) required to approximate the
coalescence dynamics in this regime whilst retaining a simple analytic expression.
%

In Figure~\ref{F:inertial_pics}, we can see that the initial dynamics of coalescence in the inertial regime is genuinely `local', as the two different geometries used for experiments (pendent drops) and simulations (free spheres) agree for the initial stages of motion, i.e.\ the deformed free surface in both the experiment and the simulation agree perfectly even though their entire shapes, by construction, do not.  It is only around $\bar{t}=0.3$ that the free surface shapes in the bridge region start to feel their entire geometry, so that the inertial regime is essentially over as the bridge's propagation is no longer `local'.  Despite this, it is seen from Figure~\ref{F:inertia_exp} that the estimates for the bridge radius stay close to the exact solution and the experimental data for a considerably longer time, remaining relatively accurate until at least $\bar{t}=0.5$ at which point $\bar{r}_b\approx 0.8$.

A shortcoming of the obtained expression is that it does not
include the influence of gravity on the drops' evolution, which
manifests itself most strongly by altering the initial shape of
drops.  This effect will be significant for drops
larger than the capillary length, which for water is of the order of
millimetres.  However, for smaller low-viscosity drops, particularly
those from around $R=10$~$\mu m$ to $R=1$~mm, where inertial effects
still dominate viscous ones, the
expression in (\ref{new_eq3}) will provide an excellent
approximation to their evolution.

\bibliographystyle{unsrt}
\bibliography{Bibliography} 

\section*{Appendix: Computation of the longitudinal curvature}

Figure~\ref{F:curvature} shows the free surface shape obtained from the computed solution for free spheres coalescing at $\textrm{Re}_i=10^3$ in the range $0.1\le\bar{r}_b\le0.7$.  Marked with crosses are the point on the free surface at which the longitudinal curvature $\bar{\kappa}_1$ changes sign, i.e.\ the crosses mark an inflection point in the free-surface profile.  The height of this inflection point $\bar{z}=\bar{z}_{inf}$ can be used to define the \emph{effective} longitudinal curvature of the bridge connecting the coalescing drops as $\bar{\kappa}_1 = 1/\bar{z}_{inf}$.  This allows us to test the assumption that $\bar{\kappa}_1 \approx 1/\bar{r}_b^2$, or alternatively that $\bar{\kappa}_1 \bar{r}_b^2 \approx 1$.
\begin{figure}[h]
     \centering
\includegraphics[scale=0.255]{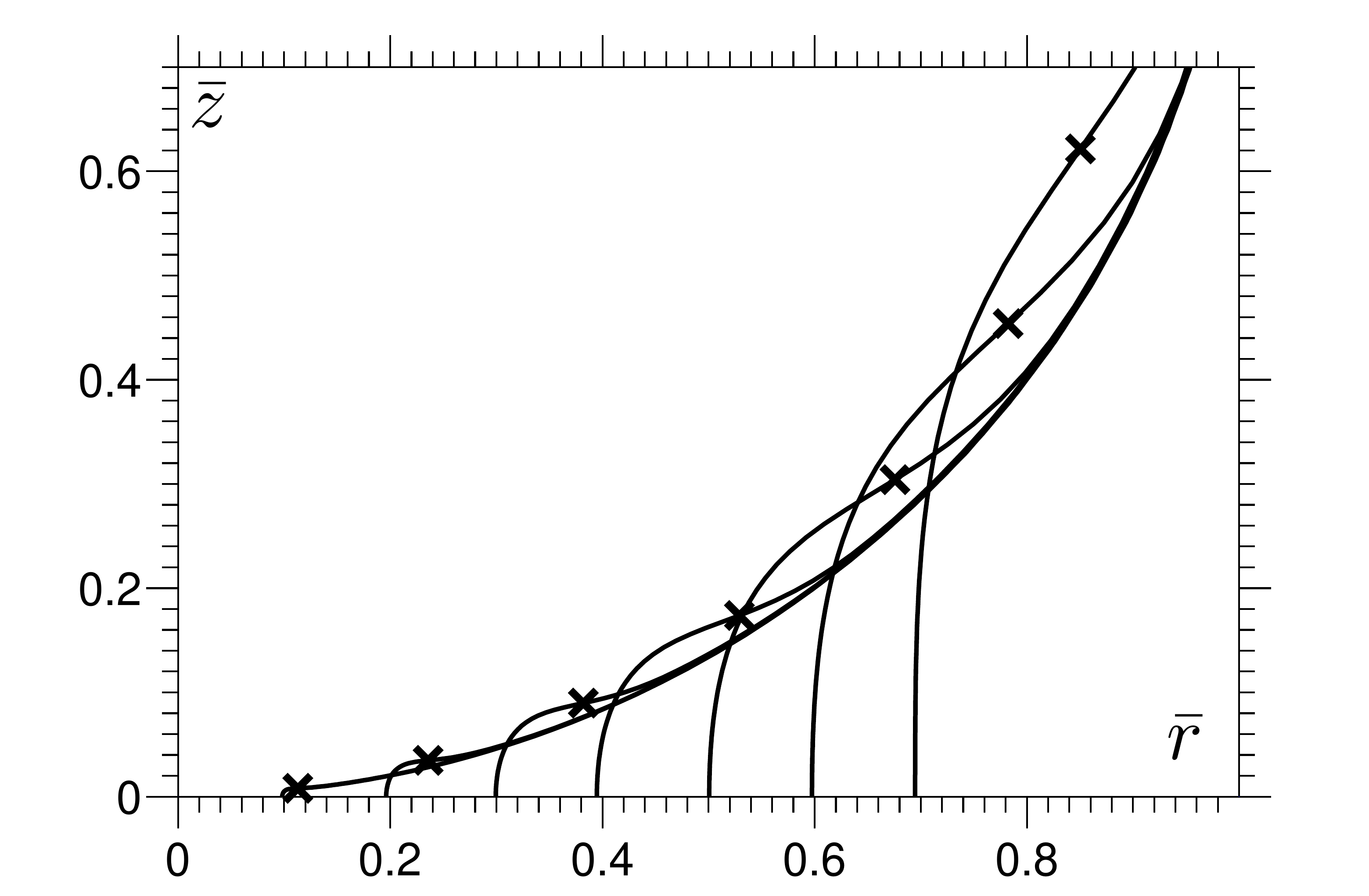}
 \caption{Snapshots of the coalescence process showing crosses at the inflection point on the free surface.}
 \label{F:shape}
\end{figure}

In Figure~\ref{F:curvature}, it can be seen that in the range $0.1\le\bar{r}_b\le0.7$ we have $\kappa_1 \bar{r}_{b}^2$ approximately constant, so that the assumed scaling behaviour sufficiently accurately reflects the exact solution: over the period considered it is in the range $\kappa_1 \bar{r}_{b}^2\in(0.8,1.2)$ so that its average value will be close to one.  Slight improvements could potentially be achieved by using a linear approximation for the curvature, but given the good agreement between the new scaling and the fully computed results this does not seem necessary.
\begin{figure}[h]
     \centering
\includegraphics[scale=0.255]{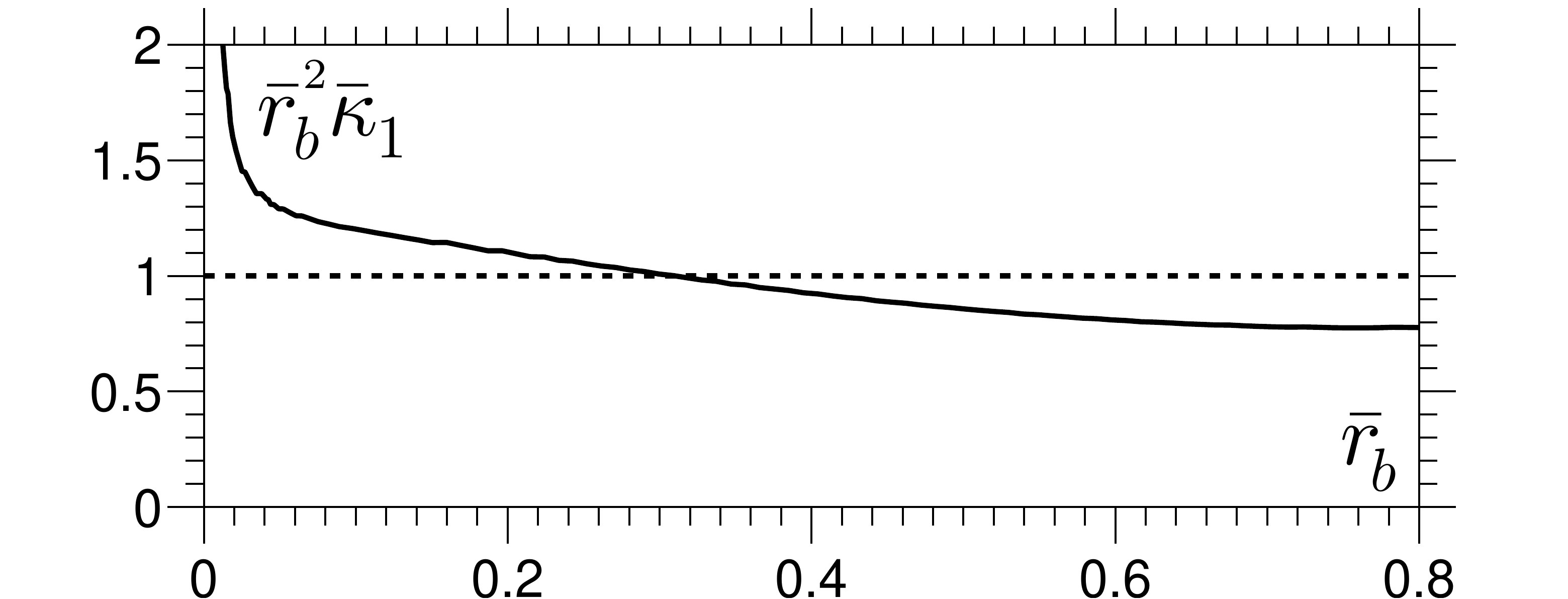}
 \caption{Evolution of the computed effective longitudinal curvature of the bridge connecting the coalescing drops $\bar{\kappa}_1$ multiplied by the square of the bridge radius $\bar{r}_b^2$.  It can be seen that $\bar{\kappa}_1\bar{r}_b^2 \approx 1$ as assumed.}
 \label{F:curvature}
\end{figure}

\end{document}